\pdfoutput=1

\documentclass[11pt]{article}

\usepackage[final]{acl}

\usepackage{times}
\usepackage{latexsym}

\usepackage[T1]{fontenc}

\usepackage[utf8]{inputenc}

\usepackage{microtype}

\usepackage{inconsolata}

\usepackage{graphicx}

\usepackage{times}
\usepackage{latexsym}
\usepackage{algorithm}
\usepackage{algpseudocode}
\usepackage{enumitem}
\usepackage[T1]{fontenc}
\usepackage[utf8]{inputenc}
\usepackage{microtype}
\usepackage{inconsolata}
\usepackage{graphicx}
\usepackage{colortbl}
\usepackage{xcolor}         
\usepackage[utf8]{inputenc} 
\usepackage[T1]{fontenc}    
\usepackage{hyperref}       
\usepackage{url}            
\usepackage{booktabs}       
\usepackage{amsfonts}       
\usepackage{nicefrac}       
\usepackage{microtype}      
\usepackage{color}
\usepackage{amsmath}
\usepackage{pifont}

\title{Language Model Based Text-to-Audio Generation: \\ Anti-Causally Aligned Collaborative Residual Transformers}

\author{
 \textbf{Juncheng Wang}$^{12}$\qquad
 \textbf{Chao Xu}$^{2}$\qquad
 \textbf{Cheng Yu}$^{2}$\qquad
 \textbf{Zhe Hu}$^{1}$\qquad
\\
 \textbf{Haoyu Xie}$^{2}$\qquad
\textbf{Guoqi Yu}$^{1}$\qquad
 \textbf{Lei Shang}$^{2}$\thanks{Correspondence}\qquad
 \textbf{Shujun Wang}$^{1}$$^*$
\\[0.7em]
 $^1$ The Hong Kong Polytechnic University\qquad 
 $^2$ Alibaba Group\qquad
}

\newcommand{\method}{\texttt{Siren}}

\begin{document}

\makeatletter
\newcommand\whline{\noalign{\ifnum0=`}\fi\hrule \@height 1.25pt \futurelet
	\reserved@a\@xhline}
\maketitle
\begin{abstract}
While language models (LMs) paired with residual vector quantization (RVQ) tokenizers have shown promise in text-to-audio (T2A) generation, they still lag behind diffusion-based models by a non-trivial margin. We identify a critical dilemma underpinning this gap: incorporating more RVQ layers improves audio reconstruction fidelity but exceeds the generation capacity of conventional LMs. To address this, we first analyze RVQ dynamics and uncover two key limitations: 1) orthogonality of features across RVQ layers hinders effective LMs training, and 2) descending semantic richness in tokens from deeper RVQ layers exacerbates exposure bias during autoregressive decoding. Based on these insights, we propose \method, a novel LM-based framework that employs multiple isolated transformers with causal conditioning and anti-causal alignment via reinforcement learning.  Extensive experiments demonstrate that \method~outperforms both existing LM-based and diffusion-based T2A systems, achieving state-of-the-art results. By bridging the representational strengths of LMs with the fidelity demands of audio synthesis, our approach repositions LMs as competitive contenders against diffusion models in T2A tasks. Moreover, by aligning audio representations with linguistic structures, \method~facilitates a promising pathway toward unified multi-modal generation frameworks. The code is released at \hyperlink{code}{https://github.com/wjc2830/Siren.git}.
\end{abstract}

\section{Introduction}
Autoregressive language models (LMs)~\citep{transformer,gpt3,llama1,palm,qwen} have emerged as the \emph{de facto} paradigm for natural language generation (NLG)~\citep{gpt3.5,gpt4,openai2022chatgpt,google2023bard,anthropic2023claude}, excelling in modeling discrete, categorical token sequences via next-token prediction. 

\begin{figure}
    \centering
    \includegraphics[width=\linewidth]{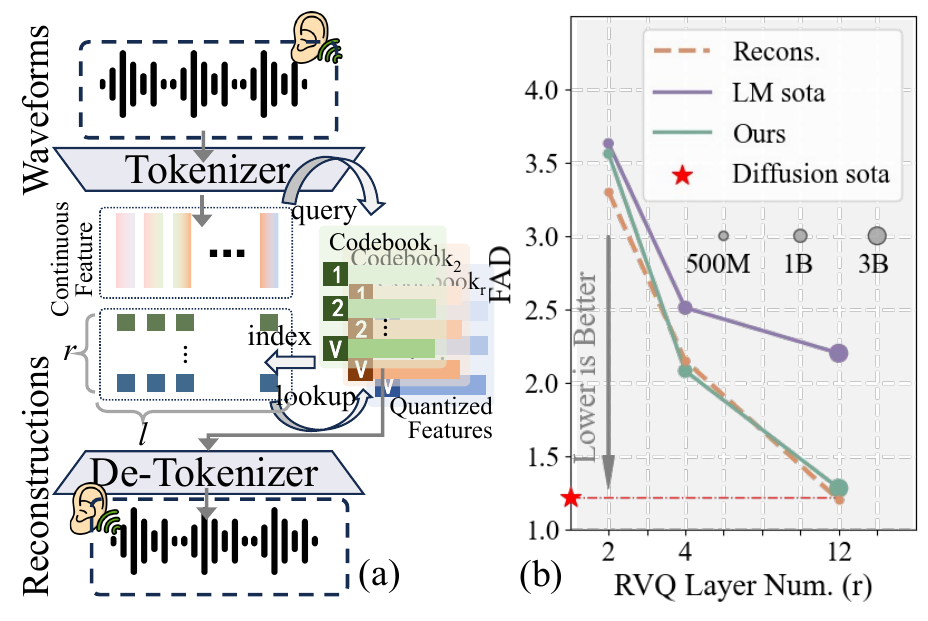}
    \vspace{-8mm}
    \caption{(a) \textbf{Residual Vector Quantization Process}, where a waveform is quantized into $r\times l$ discrete tokens. With $r=1$, it degenerates to naive VQ. (b) \textbf{Performance changing curves as $r$ going larger.} }
    \vspace{-6mm}
    \label{fig-teaser}
\end{figure}

Inspired by the advancements of LMs in NLG, recent works have adopted the LMs for text-to-audio (T2A) generation~\citep{liu2024audioldm,ziv2024masked,copet2024simple}. However, this application presents challenges due to the representational gap between discrete LM tokens and continuous audio waveforms.

To bridge this gap, vector quantization (VQ)~\citep{vqvae,vqvae2} becomes essential to map continuous audio into discrete token sequences, enabling audio generation through \emph{next-token prediction}. 

Nevertheless, considering VQ suffers from lossy compression~\citep{vqgan,vit-vqgan,magvit2}, \emph{residual vector quantization} (RVQ) has been widely adopted~\citep{wu2019vector,encodec,dac} (Figure~\ref{fig-teaser}-a), where each temporal audio slice is recursively quantized into $r>1$ residual tokens. Unlike standard VQ, RVQ necessitates \emph{next-r-token prediction}~\citep{rqtransformer,copet2024simple,audiogen}.

Despite these advances, LM-based T2A systems~\citep{audiogen,copet2024simple,ziv2024masked} still lag significantly behind diffusion models~\citep{xue2024auffusion,MMAudio,audioX,GenAU}. For instance, on the AudioCaps benchmark~\citep{kim2019audiocaps}, state-of-the-art LMs~\citep{copet2024simple} trail diffusion baselines~\citep{GenAU} by 45\% in Fréchet Audio Distance (FAD: 2.20 vs. 1.22; Figure~\ref{fig-teaser}-b). This gap persists even as RVQ tokenizers achieve high reconstruction fidelity with deeper layers ($r^\uparrow$), revealing a critical \textbf{dilemma}: while increasing $r$ improves tokenizer reconstruction quality, it overtaxes the generative capacity of LMs. 
In Figure~\ref{fig-teaser}-b, LM-based model struggles to effectively use tokenizer with deeper RVQ layers. This leads to a performance ceiling, where increasing $r$ does not lead to improvements in generation quality, despite enhanced reconstruction capability.

This raises a critical question for high-fidelity audio generation: \emph{how can LMs more effectively predict next-$r$-codes as $r$ increases?}
Our pilot studies of RVQ properties and their impact on LMs reveals two key challenges.

\underline{(1) Feature Orthogonality}: Quantized features from distinct RVQ layers exhibit near-orthogonality in latent space. Standard approaches~\citep{audiogen,copet2024simple} typically employ a shared transformer to predict all RVQ tokens, which forces the model to aggregate diverse gradients from orthogonal feature targets. Nevertheless, this gradient conflict impedes model convergence and limits expressiveness;
\underline{(2) Semantic Degradation in Deeper RVQ Layers}: As RVQ layer goes deeper, the semantic richness within the corresponding quantized features decreases, resulting in a learning difficulty imbalance across different RVQ codebook classifiers. 
Consequently, such imbalance leads to different fitting degree to each RVQ codes.
During autoregressive decoding, this imbalance exacerbates the issue of \emph{exposure bias}~\citep{schmidt2019generalization}, where errors accumulate as the model transitions from the ground-truth-conditioned training to self-conditioned inference.

Inspired by these findings, we propose \method, anti-cau\textbf{S}ally al\textbf{I}gned collaborative \textbf{RE}sidual tra\textbf{N}sformers.
To address the first challenge of gradient conflicts caused by orthogonal RVQ features, \method~ distributes the prediction of $r$ RVQ codes across $r/2$ collaborative transformers, which are trained independently, thus reducing learning diversity within each transformer. To preserve the causal dependencies among RVQ codes, accumulated conditions across models are further introduced to establish collaboration between transformers.

However, this partitioning alone does not address the inherent imbalance in learning difficulty across RVQ codes. The first transformer—assigned to semantic-rich shallow codes—faces high stochasticity in its predictions~\citep{Improving25}, propagating unstable conditions to downstream models and thereby amplifies exposure bias, as described in the aforementioned second challenge of semantic degradation.
To mitigate this, we further fine-tune the first transformer using reinforcement learning (RL), to align the outputs from the first transformer towards the conditional \emph{preferences} of subsequent transformers, enhancing decoding stability and fidelity of generated audios.

We conduct extensive experiments and empirically demonstrate that \method~ achieves the state-of-the-art (\emph{SOTA}) performance comparing with both LM-based and diffusion-based T2A methods.
By reconciling the strengths of LMs with RVQ dynamics, our approach bridges the gap between discrete and continuous generation paradigms, paving the way for unified multi-modal generation where discrete tokens serve as a universal interface.
In summary, our contributions are three-fold:
\begin{itemize}[noitemsep,nolistsep,wide]
    \item We identify two critical limitations of RVQ in LM-based generation: RVQ layer-wise orthogonality of quantized features and exposure bias due to semantic degradation in deeper layers.
    \item We propose \method, a novel LM architecture that employ collaborative transformers with RL based anti-causal alignment, to resolve gradient conflicts and distribution drift.
    \item We demonstrate \method's \emph{SOTA} performance through extensive experiments, showing that it outperforms both LMs and diffusion based models.
\end{itemize}

\vspace{-2mm}
\section{Preliminaries}
\vspace{-1mm}
\noindent\textbf{Residual Vector Quantization for High Quality Reconstruction}
Consider an audio waveform $x\in\mathbb{R}^{l_{wav}\times C_{wav}}$, with $C_{wav}$ as channel number, $l_{wav}$ as duration.
To apply Language Models (LMs) to audios generation via next-token prediction, residual vector quantization (RVQ)~\citep{dac,wavtokenizer} tokenizes $x$ into $f=\mathcal{E}(x)\in \mathbb{R}^{l\times C}$ feature through a tokenizer $\mathcal{E}$, with $l$ as the compressed temporal length, $C$ the feature channel. 
With the continuous feature, it is further quantized into discrete codes $q\in[V]^{r\times l}$, where $V$ is the length of codebook, namely \emph{vocabulary}, $r$ is the number of RVQ layer.
Specifically, each temporal step continual feature $f_t,t\le l$ is decomposed of $r$ layer features in a residual way:
\vspace{-4mm}
\begin{equation}
\begin{aligned}
\label{eq:RVQ}
    &f_t^0=f_t,\ f_t^j=f_t^{j-1}-\hat{f}_t^{j-1},\ q_t^j=\mathcal{Q}^j(f_t^{j-1}),
    \\
    &\hat{f}_t^j=\mathtt{lookup}(Z^j,q_t^j),\ \hat{x}_t=\mathcal{D}\textcolor{black}{(} 
    \sum_{j=1}^{j\le r}\hat{f}_t^j
    \textcolor{black}{)},
\end{aligned}
\end{equation}
where $\mathcal{Q}^j$ is an independent quantizer\footnote{See appendix for detailed process of quantization.} for $j^{th}$ layer that maintains an independent codebook $Z^j\in\mathbb{R}^{V\times C}$ containing $V$ vectors, $\mathtt{lookup}(Z, q)$ means finding $q^{th}$ vector in codebook $Z$. 
With original audio feature decomposed and quantized individually into $r\times l$ discrete tokens, the reconstructed audio can be derived from the last term in Eq.~\ref{eq:RVQ} through a de-tokenizer $\mathcal{D}$.
During tokenizer training phase, it is optimized through a general audio reconstruction goal with different loss designs.
Empirically, one can derive better reconstruction with larger $r$.

\paragraph{Enhanced Generation Complexity of RVQ for Transformer Generation}
With an audio tokenized into above tokens, the mainstream LM-based generators, like AudioGen~\cite{audiogen}, have a unified transformer model $\mathcal{F}$ to conduct the \emph{next-time hidden state prediction}, which will be fed into $r$ independent classifier heads $\{\mathcal{C}_j\}_{j=1}^{r}$ to conduct $r$ times $V$-way classifications to finish \emph{next-r-code prediction}.
Then, the predicted $r$ codes are transferred into corresponding $r$ embeddings, and summed over $r$ to derive input embedding to condition the further next time prediction.
According to the above pipeline of generating audios from discrete tokens,
decomposing one temporal step feature into $r$ discrete tokens introduces better reconstruction fidelity, but the complexity of transformer generation that conducts $r$ times $V$-way classifications at each time step is also enhanced.
Despite previous works~\citep{audiogen,copet2024simple,ziv2024masked} choose a relatively small $r$, like $r=4$, its overall performance lags behind mainstreams, due to being unable to generate audios with $r$ scaled up. 
Thus, it is critical to answer \textbf{how can we better predict multi-RVQ codes when $r$ is large} to derive high fidelity audio generation.

\section{Dilemma of Reconstruction Quality and Generation Hardness: A Challenge}
\begin{figure}[t]
    \centering
    \includegraphics[width=1\linewidth]{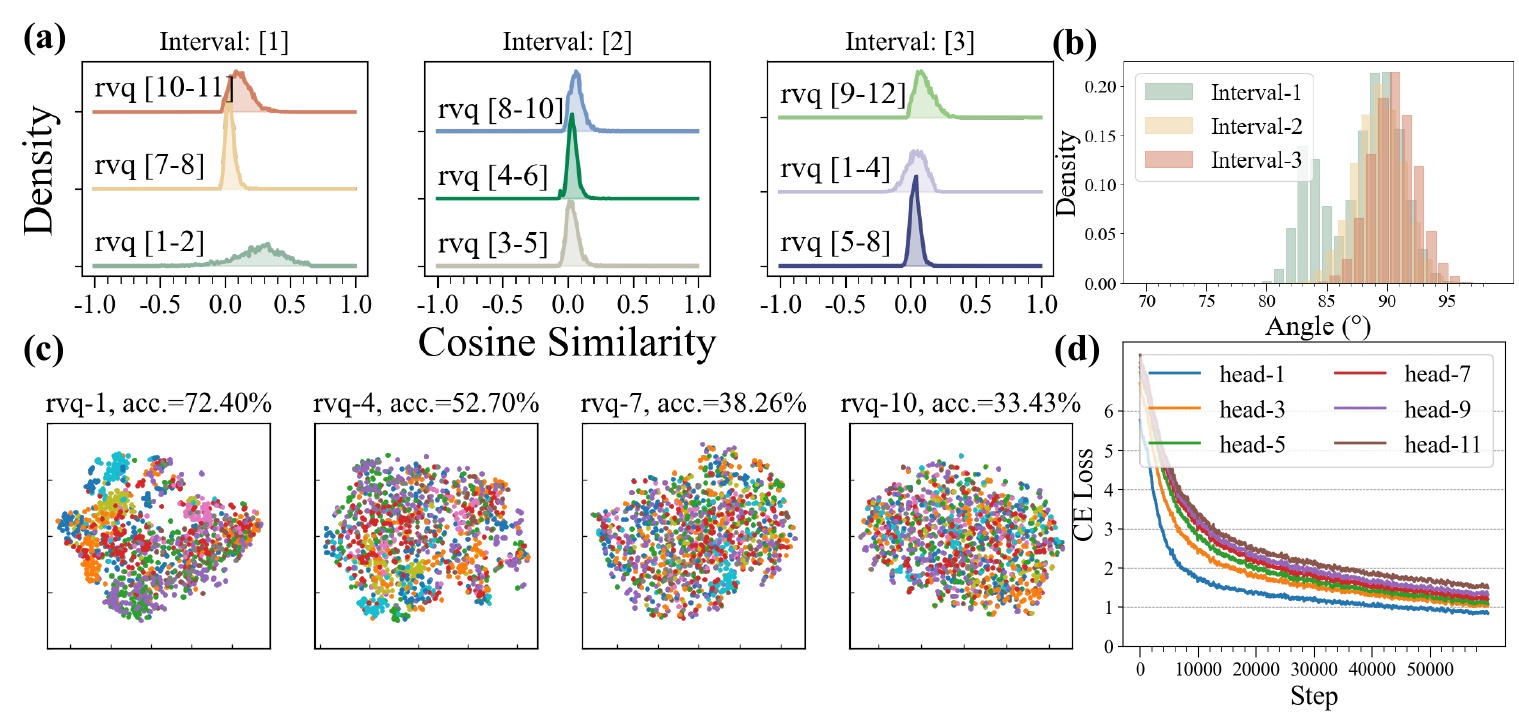}
     \vspace{-8mm}
    \caption{(a) Cosine distributions between quantized features from different RVQ layer. The \emph{Interval} here means the layer interval between two paired features. (b) The angle distribution between gradient vectors that different classifier heads backpropagated to the shared transformer. (c) T-SNE processed quantized features from different RVQ layer, where each audio class has been colored, and corresponding accuracy is on the top. (d) Convergence curves of learning different RVQ codes. \emph{Due to page limitation, we have placed the whole layer results in the Appendix.} Please zoom in for more details.}
    \vspace{-5mm}
    \label{fig-pilot_study}
\end{figure}
To better understand RVQ, we conduct several pilot studies to reveal two key properties, and their corresponding influences to LMs training.
\paragraph{Property 1:}\emph{Orthogonality between two quantized features from different residual layers.}
Formally, given $j^{th}$ RVQ layer quantized feature $\hat{f}^j$, it attempts to model the residual difference between continuous feature with quantized one.
Thus, the knowledge represented by each residual layer usually is distinct from others.
To verify this, we utilize a trained RVQ tokenizer~\citep{dac} with $r=12$, and visualize the distribution of cosine similarity between any two quantized features from different residual layers ($j_1$ and $j_2$) of the same time-step $t$ on AudioCaps~\citep{kim2019audiocaps}, like $\mathtt{cosine}(\hat{f}_t^{j_1},\hat{f}_t^{j_2})$. As shown by Figure~\ref{fig-pilot_study}-(a), the cosine similarities are distributed around 0, which represents (approximately) orthogonality.

\paragraph{Influence 1:}\emph{Orthogonality introduces diverse learning direction of transformers.}
With RVQ decomposing a continuous feature into $r$ (approximately) orthogonal discrete code features, the LM is required to fit $r$ different distributions
of $\{p(q_t^j|q_{<t},c)\}_{j=1}^r$ 
simultaneously, where $c$ is the textual prompts.
Considering the diverse distributions, it incurs diverse learning directions to transformer, which significantly enhances the learning hardness.
To demonstrate it, we utilize the LM based AudioGen to predict aforementioned RVQ tokenizer with $r=12$.
Specifically, we visualize the gradients that each independent classifier backpropagates to the last 
layer of $\mathcal{F}$.
\begin{figure*}
    \centering
    \includegraphics[width=0.9\linewidth]{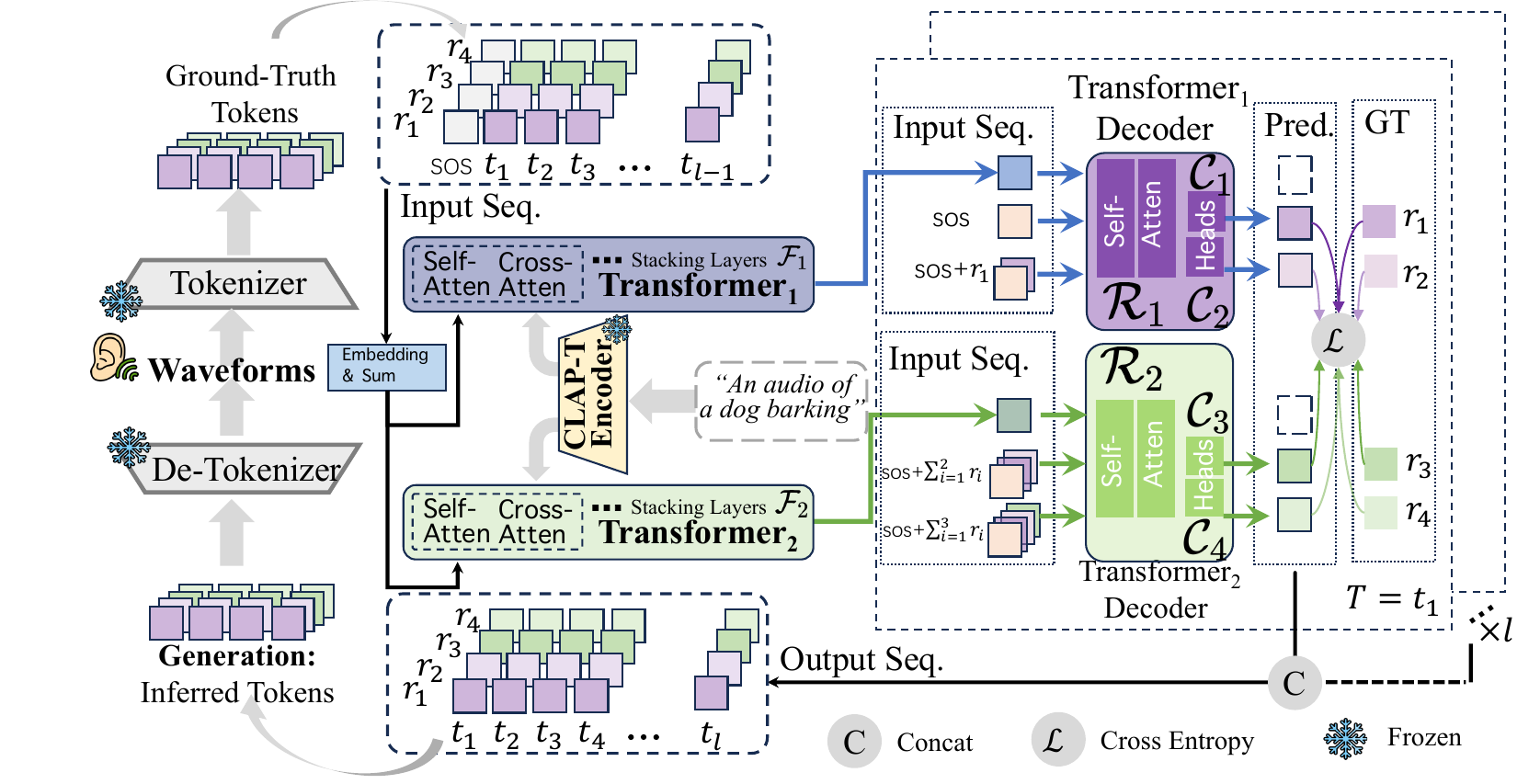}
    \caption{The pipeline of the proposed \method. Due to spatial restriction, we simplify our real setting ($r=12$) into a case that $r=4$ where $r/2=2$ parallel transformers are fed with the same input to conduct \emph{next-time feature prediction}, which are then factorized into corresponding two adjacent RVQ codes through independent decoders and classifiers. Then, the predicted tokens can be concatenated together and fed into de-tokenizer to recover into audio waveform.
    In training phase, teacher forcing is adopted where the contextual tokens are all the ground-truth.}
    \label{fig-main_framework}
    \vspace{-0.2cm}
\end{figure*}
\noindent As shown by Figure~\ref{fig-pilot_study}-(b), the gradient directions of each classification heads are different. When aggregating the diverse gradients to the shared transformer, it increases the difficulty of convergence.

\paragraph{Property 2:} \emph{Descending semantic richness when RVQ layer goes deeper.}
We further investigate how each residual feature contributes to the final reconstruction.
To this end, we first extract temporally complete discrete codes of an audio from different residual layers, \emph{i.e.,} $\{\hat{f}^j\in\mathbb{R}^{l\times C}\}^{r}_{j=1}$. For $j^{th}$ layer feature, on AVSync-15 dataset~\citep{avsync}, which has audios from $15$ categories, we utilize these embeddings to train $r$ audio classifiers.
Following above setting, we utilize the same RVQ tokenizer with $r=12$. Then, we exhibit classification results in Figure~\ref{fig-pilot_study}-(c).
It is shown that as RVQ layer goes deeper, the classification performance goes worse, meaning less semantic richness. 

\paragraph{Influence 2:} \emph{Imbalanced learning hardness among different residual layer code prediction heads.}
As aforementioned, as RVQ layer $j$ being larger, the quantized feature $\hat{f}^j$ contains less semantics. This makes $p(q_t^j|q_{<t})$ harder to fit.
To support this, for above trained AudioGen, we visualize their loss curves.
It is evident in Figure~\ref{fig-pilot_study}-(d), the convergence becomes slower and worse as layer goes deeper.
In autoregressive training, the model is conditioned by ground-truth tokens from previous temporal steps or RVQ layers.
While in inference, it is conditioned by self-inferred tokens. 
Thus, such an imbalanced convergence degree results in worse inconsistency over conditions, which reveals exposure bias.

\noindent
\textbf{In sum}, the key insights from our pilot studies can be concluded as follows: 
1) the orthogonality between RVQ layer features makes one model hard to fit them simultaneously; 
2) the imbalanced semantic richness among RVQ layers results in exposure bias among RVQ layers.
These hinder the high quality audio generation when $r$ is large.

\section{Siren: Anti-Causally Aligned Collaborative Residual Transformers}
Based on the above insights, we propose \method, anti-cau\textbf{S}ally al\textbf{I}gned collaborative \textbf{RE}sidual tra\textbf{N}sformers, with pipelines presented in Figure~\ref{fig-main_framework}\&~\ref{fig-RL_pipeline}, algorithms in Appendix~\ref{app:algorithms}. 
Firstly, to alleviate the diverse gradient directions among learning different RVQ layers aforementioned in \emph{Influence 1},
\method~distributes $r$ RVQ codes prediction into $r/2$ collaborative transformers, which are trained independently, thus reducing diversity within each transformer, and making optimization easier. 
To further maintain the causal relationship between codes, accumulated conditions across models are used to establish collaboration between transformers. 
Secondly, individually training transformers cannot mitigate exposure bias incurred by different fitting hardness, but makes it worse. 
To address it, we deem the first model that is responsible for $1^{st}$ and $2^{nd}$ RVQ codes with most rich semantics, faces high stochasticity during sampling, propagating unstable conditions to downstream models.
Thus, we introduce reinforcement learning (RL) to align the sampled output from the first model to the condition preferences of other models with an anti-causal direction, aiming to enhance the overall performance of collaborative transformers.

\subsection{Collaborative Residual Transformers}
\paragraph{Isolation for Better Optimization} 
To remedy the above discussed diverse gradient directions to the shared transformer, we distribute $r$ codes prediction tasks to $K$ models, each of which is only responsible for its assigned codes. Concretely, $k^{th}$ model $\mathcal{H}_k$ is individually trained to predict $2k^{th}$ and $(2k+1)^{th}$ layer codes to isolate diverse optimization directions.
Given RVQ sets $\{(q_t^1,...,q_t^r)\}_{t=1}^{l}$, $k^{th}$ model is to fit the distribution of:
\begin{equation}
\begin{aligned}
    \label{}
    &p_{\theta_k}(q_1^{k_x},q_2^{k_x},...,q_l^{k_x})=\prod_{t=1}^{l} p_{\theta_{k}}(q_t^{k_x}|\hat{q}_{<t}^{1},...\hat{q}_{<t}^{r}),
\end{aligned}
\end{equation}
where $k_x=2k\text{ or }2k+1$, $\theta_k$ denotes the independently maintained parameters of transformers and classifier heads. Hence, compared to having a single model predict the diverse distributions, using $K$ models makes the task more manageable.
\paragraph{Collaboration for Causal Relationship}
Although isolating the training of each model facilitates better learning target concentration, it incurs sampling cooperation issue during generation.
Specifically, let $h_{k,t}=\mathcal{F}_k(q_{<t}^1,...,q_{<t}^r,c)$ be the predicted next-time step feature from backbone transformer, the generation of code is to sample from the probability over codebook predicted by corresponding transformer, like:
\begin{equation}
    \label{eq:issue}
    p_{\theta_k}(q^{k_x}_{t})=\text{softmax}\textcolor{red}{(}
        \mathcal{C}_{k_x}(h_{k,t})
    \textcolor{red}{)}.
\end{equation}
When transformers are independent, the generation among codes from the same temporal step but different residual layers are totally isolated. For example, in Eq.~\ref{eq:issue}, given $k_x>1$, the information of $q_t^{<k_x}$ is agnostic when predicting $q_t^{k_x}$.
But the tokenizer training phase leaves a causal relationship among codes, \emph{i.e.}, it depends on $q_t^{<k_x}$ to derive $q_t^{k_x}$.
Thus, the independent prediction in Eq.~\ref{eq:issue} damages the causality and hinders generation.

To remedy this, additional transformer decoder layers $\mathcal{R}_k$ are introduced to process the causal relationship among residual codes.
Concretely,  
$\mathcal{R}_k$ predicts $(k_x)^{th}$ code conditioned by previous residual codes from the same temporal step, like:
\begin{equation}
    p_{\theta_k}(q^{k_x}_t)=\text{softmax}\textcolor{red}{(}
        \mathcal{C}_{k_x}\textcolor{cyan}{(}
            \mathcal{R}_k(h_{k,t},q_t^{<k_x})
        \textcolor{cyan}{)}
    \textcolor{red}{)}.
\end{equation}

\begin{figure}
    \centering
    \includegraphics[width=0.9\linewidth]{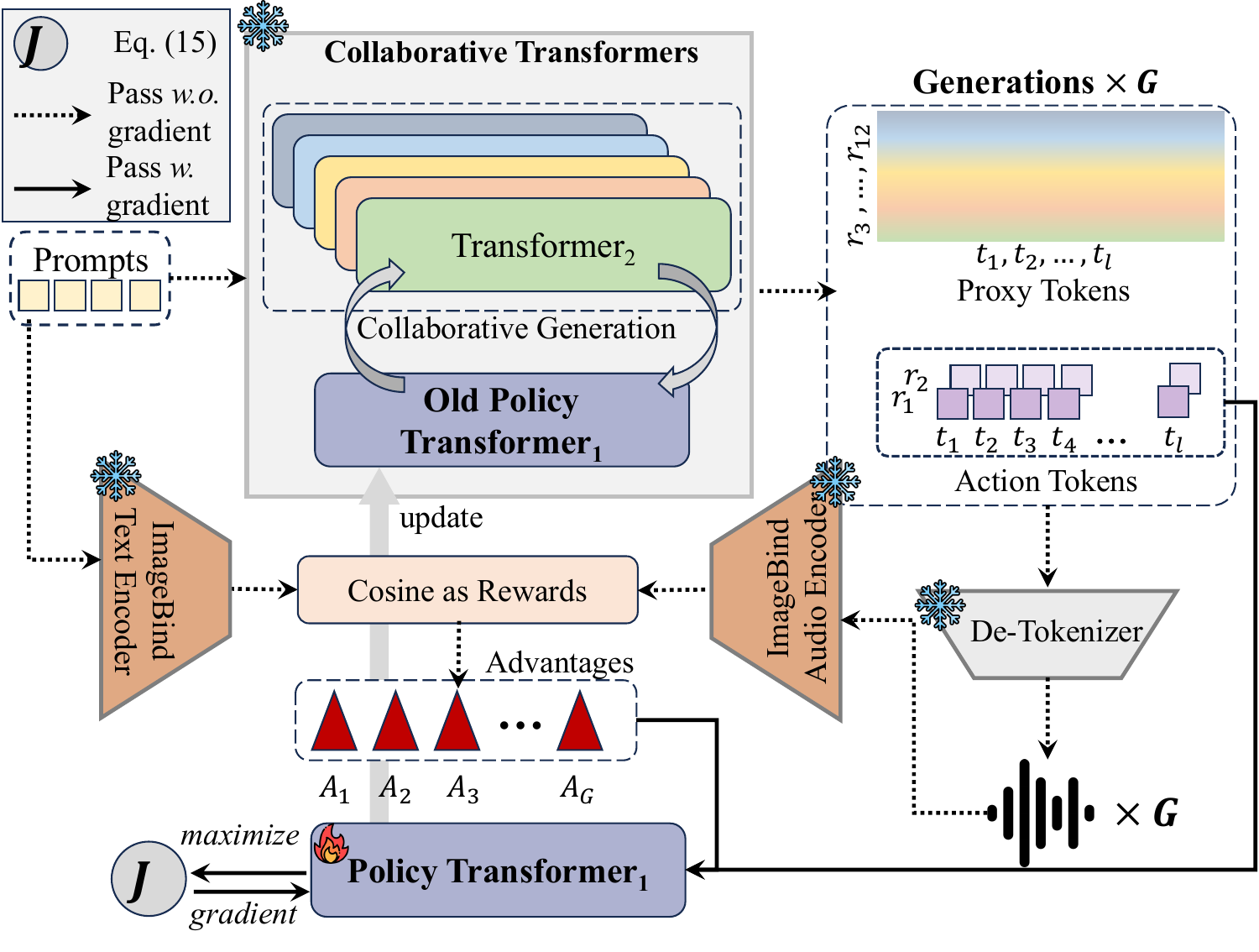}
    \caption{Overall pipeline for reinforcement learning. Given a prompt, first stage trained models generate whole audio tokens, where the output of the first transformer is used as action to align with the conditional preference of other models by an audio quality reward.}
    \label{fig-RL_pipeline}
    \vspace{-0.5cm}
\end{figure}
In $k^{th}$ model, two adjacent $2k^{th}$ and $(2k+1)^{th}$ codes facilitates modeling the residual context.
To shorten the context length of $\mathcal{R}_k$ while keeping the information flow from $1^{st}$ to $(2k-1)^{th}$ residual layers, we fold codes predicted from other transformers by accumulation, like:
\begin{equation}
\begin{aligned}    &\mathcal{R}_k([\cdot,q_t^{2k},q_t^{2k+1}])=\\
&\mathcal{R}_k([h_{k,t},\texttt{sos}+\sum_{j=1}^{2k-1}\hat{f}_t^j,\texttt{sos}+\sum_{j=1}^{2k}\hat{f}_t^j]), \\
&\text{where } \hat{f}_t^j=\texttt{lookup}(Z^j,q^j_t),
\end{aligned}
\end{equation}
This equation represents the input and output sequence of transformer decoder, with a causal inverted triangular attention mask. $\texttt{sos}$ is a learnable start token. The training target for $k^{th}$ model is:
\begin{equation}
    \label{eq:sup_loss}
    \begin{aligned}
    &\mathcal{L}_k = \frac{1}{l}\sum_{t=1}^{l}(
            \mathcal{L}_{ce}(z^{2k}_t,q^{2k}_t)+\mathcal{L}_{ce}(z^{2k+1}_t,q^{2k+1}_t)
            ), \\
            &\text{where } z^{k_x}_t=\mathcal{C}_{k_x}\textcolor{red}{(}
            \mathcal{R}_k\textcolor{cyan}{(}
                \mathcal{F}_k(q^{1}_{<t},...,q^{r}_{<t},c),q_t^{<k_x}
                \textcolor{cyan}{)}
            \textcolor{red}{)}.
    \end{aligned}
\end{equation}

\subsection{Anti-Causally Alignment in RL}
With individually trained transformers, the exposure bias can be worse, due to imbalance described in \emph{Property 2} and \emph{Influence 2}.

To alleviate this, \method~ presents a second stage training by aligning output from the first model, \emph{i.e.,} the one for $1^{st}$ and $2^{nd}$ codes, to others model's preferences, using reinforcement learning (Figure~\ref{fig-RL_pipeline}). 
Considering its difference with causal order of residual codes, 
we name it as anti-causal alignment, \emph{i.e.,} the conditional preference of later models are known, and set as anchors to be aligned. We adopt such a direction because: 1) As presented in \emph{Property 2}, shallow RVQ layers contain more semantic information, leading to more stochastic sampling results~\citep{Improving25}; 2) According to \emph{Influence 2}, learning deeper codes is harder, so fine-tuning the first transformer would be more accessible; 3) In autoregressive generation, initial tokens play a key role in determining the overall semantics of the sequence~\citep{barbero2025llms}.

Concretely, we define a Markov Decision Process (MDP)~\citep{gpt3.5} of tokens output from first transformer, where $1^{st}$ and $2^{nd}$ RVQ codes as the action sequence and tokens from other transformers as proxy tokens for reward.
For each temporal step, given a state $\mathbf{s}_t=(\hat{q}_{<t}^{1},...,\hat{q}_{<t}^{r},c)$, the first transformer generates action $\mathbf{q}_t = (\hat{q}^1_t, \hat{q}^2_t)$ with a policy $\pi(\cdot|\mathbf{s}_t)$, defined as:
\begin{equation}
\pi(\cdot|\mathbf{s}_t)=\texttt{softmax}\textcolor{red}{(}
        \mathcal{H}_1\textcolor{blue}{(}
            \hat{q}_{<t}^{1},...,\hat{q}_{<t}^{r},c
        \textcolor{blue}{)}
    \textcolor{red}{)},
\end{equation}
where $\mathcal{H}_1=\mathcal{C}_1\times \mathcal{R}_1 \times \mathcal{F}_1$.

Then, we need to compute the reward for this generation. 
Recap our target is to align this action $\mathbf{q}\in[V]^{2\times l}$ to the preference of other transformers, yet we cannot directly derive quantitative metric to this \emph{preference}, thus we use the finally generated audio quality as a proxy reward. 
Technically, we introduce the de-tokenizer $\mathcal{D}$ and ImageBind model~\citep{girdhar2023imagebind} $\phi_{text}$ and $\phi_{audio}$ to compute the reward as:
\vspace{-2mm}
\begin{equation}
    \label{eq:reward}
    R=\text{cosine}\textcolor{blue}{(}
        \phi_{audio}\textcolor{red}{(}
            \mathcal{D}\textcolor{green}{(}
                \text{cat}(\mathbf{q},\mathbf{q}_{prox})
            \textcolor{green}{)}
        \textcolor{red}{)}, 
        \phi_{text}\textcolor{cyan}{(}c\textcolor{cyan}{)}
    \textcolor{blue}{)},
\end{equation}
where $\mathbf{q}_{prox}\in[V]^{(r-2)\times l}$ are the proxy tokens. 

Later, we use the proximal policy optimization (PPO)~\cite{PPO} with a value-model free modification~\citep{GRPO,DAPO,VAPO,CPPO} to seek efficiency.
For a specific prompt-generated audio pair, 
the policy $\pi_{\theta_{old}}$ samples a group of individual action sequences $\{\mathbf{q}_i\}_{i=1}^{G}$. Then, the advantage of the $i^{th}$ action sequence is calculated by normalizing the group-wise rewards $\{R_i\}^G_{i=1}$:
\begin{equation}
    \label{eq:advantage}
    A_i=\frac{R_i-\text{mean}(\{R_i\}^G_{i=1})}{\text{std}(\{R_i\}^G_{i=1})}.
\end{equation}
\vspace{-1mm}
The training objective $\mathcal{J}(\theta_1)$ is maximized as:
\begin{equation}
\label{eq:RLObject}
\begin{aligned}
&\mathbb{E}_{\{(q_t^1)_i\}_{i=1}^{G} \sim \pi_{\theta_{old}}(\cdot|\mathbf{q}_{<t},c), 
\{(q_t^2)_i\}_{i=1}^{G} \sim \pi_{\theta_{old}}(\cdot|\mathbf{q}_{<t},c,q_t^1)}\\
&\frac{1}{G}\sum_{i=1}^{G}\frac{1}{l}\sum_{t=1}^{l}\frac{1}{2}\sum_{j=1}^{2}
\min (r_{ij}A_i,v_{ij}A_i), \text{s.t. } |A_i|\ge\gamma , 
\end{aligned}
\end{equation}
\vspace{-3mm}
\begin{equation}
\begin{aligned}
&\text{where } v_{ij}=\text{clip}(r_{ij},1-\epsilon_d,1+\epsilon _u),\\
&\quad\quad\ \ \  r_{ij}=\frac{\pi_{\theta _1}((q_t^j)_i|(\mathbf{q}_{<t})_i,c,(q_t^{j-1})_i)}{\pi_{\theta _{old}}((q_t^j)_i|(\mathbf{q}_{<t})_i,c,(q_t^{j-1})_i)},
\end{aligned}
\end{equation}
where $\epsilon_d$ and $\epsilon_u$ are two coefficients to restrict the update the model in each step, and $\gamma$ is a threshold to filter out those action sequences with lower absolute rewards to make model concentrate on sequences with more extinguish rewards.

After fine-tuning the first transformer with above reinforcement objective to align the condition preference of other transformers, it alleviates the exposure bias issue to enhance generation quality.

\section{Experiment}
\subsection{Settings}
\paragraph{Dataset and Training Details.} Following previous works~\citep{xue2024auffusion,audiogen}, we collect audios from AudioSet~\citep{audioset}, AudioCaps~\citep{kim2019audiocaps}, Clotho~\citep{clotho}, ESC50~\citep{esc50}, FreeSound~\citep{freesound}, VGGSound~\citep{chen2020vggsound}, AVSync-15~\cite{avsync}, BBC Sound Effects~\footnote{https://sound-effects.bbcrewind.co.uk/}, and SoundBible~\footnote{https://soundbible.com/}. We utilize QWen-2 Audio~\citep{chu2024qwen2} to annotate those audios without textual prompts, which are polished through QWen-2.5~\citep{qwen}.
Then, we filter the text-audio pairs by CLAP score~\citep{CLAP} with a threshold.
In this paper, we utilize three levels of thresholds to obtain data collects for $1.6B$, $3.1B$, and reinforcement training. See appendix for detailed data engine process.
As for training, we utilize a RVQ tokenizer~\citep{dac} with $r=12$.
Our first stage training is conducted through $150K$, and $500K$ steps for two model size variants.
Then, the first transformer is further fine-tuned by additional $2K$ steps.
The trained models are then tested on AudioCaps dataset.
\paragraph{Metrics}
Following previous works~\citep{liu2023audioldm}, we employ Fréchet Distance (FD), Fréchet Audio Distance (FAD), Kullback-Leibler (KL) 
\begin{table*}[]
\centering

\renewcommand{\arraystretch}{1.15}{
\resizebox{\textwidth}{!}{
\begin{tabular}{ccccccccc}
\whline
\multicolumn{1}{c|}{Method}        & \multicolumn{1}{c|}{\#Parameter} & \multicolumn{1}{c|}{\#Train Sample} & \multicolumn{1}{c|}{\# Gen.Time} & \multicolumn{1}{c|}{FAD$^\downarrow$}  & \multicolumn{1}{c|}{FD$^\downarrow$}    & \multicolumn{1}{c|}{ISC$^\uparrow$}   & \multicolumn{1}{c|}{KL$^\downarrow$}   & CLAP$^\uparrow$  \\ \whline
\multicolumn{9}{c}{continuous tokens (diffusion based)}                                                                                                                                                                                                                  \\ \hline
\multicolumn{1}{c|}{AudioLDM2~\citep{liu2024audioldm}}     & \multicolumn{1}{c|}{712M}        & \multicolumn{1}{c|}{760K}           & \multicolumn{1}{c|}{53s}         & \multicolumn{1}{c|}{1.82} & \multicolumn{1}{c|}{31.02} & \multicolumn{1}{c|}{8.46}  & \multicolumn{1}{c|}{1.69} & 15.73 \\ \hline
\multicolumn{1}{c|}{Auffusion~\citep{xue2024auffusion}}     & \multicolumn{1}{c|}{1.1B}        & \multicolumn{1}{c|}{470K}           & \multicolumn{1}{c|}{58s}         & \multicolumn{1}{c|}{2.22} & \multicolumn{1}{c|}{18.73} & \multicolumn{1}{c|}{12.73} & \multicolumn{1}{c|}{1.39} & 21.35 \\ \hline
\multicolumn{1}{c|}{TANGO-Flux~\citep{hung2024tangoflux}}    & \multicolumn{1}{c|}{515M}        & \multicolumn{1}{c|}{445K}           & \multicolumn{1}{c|}{16s}         & \multicolumn{1}{c|}{2.25} & \multicolumn{1}{c|}{17.99} & \multicolumn{1}{c|}{12.81} & \multicolumn{1}{c|}{1.41} & 24.65 \\ \hline
\multicolumn{1}{c|}{ETTA~\citep{ETTA}}        & \multicolumn{1}{c|}{-}        & \multicolumn{1}{c|}{2.77M}   & \multicolumn{1}{c|}{-}         & \multicolumn{1}{c|}{1.89} & \multicolumn{1}{c|}{11.13} & \multicolumn{1}{c|}{\textbf{15.05}} & \multicolumn{1}{c|}{\textbf{1.26}} & - \\ \hline
\multicolumn{1}{c|}{GenAU~\citep{GenAU}}         & \multicolumn{1}{c|}{1.25B}       & \multicolumn{1}{c|}{811K}           & \multicolumn{1}{c|}{52s}         & \multicolumn{1}{c|}{\textbf{1.22}} & \multicolumn{1}{c|}{15.86} & \multicolumn{1}{c|}{11.90} & \multicolumn{1}{c|}{\underline{1.28}} & 24.07 \\ \hline
\multicolumn{1}{c|}{MMAudio~\citep{MMAudio}}       & \multicolumn{1}{c|}{1.03B}       & \multicolumn{1}{c|}{951K}           & \multicolumn{1}{c|}{7s}          & \multicolumn{1}{c|}{4.21} & \multicolumn{1}{c|}{13.63} & \multicolumn{1}{c|}{12.45} & \multicolumn{1}{c|}{1.40} & \textbf{30.63} \\ \hline
\multicolumn{1}{c|}{AudioX~\citep{audioX}}        & \multicolumn{1}{c|}{1.1B}        & \multicolumn{1}{c|}{330K (+5.9M)}   & \multicolumn{1}{c|}{58s}         & \multicolumn{1}{c|}{1.63} & \multicolumn{1}{c|}{11.67} & \multicolumn{1}{c|}{12.44} & \multicolumn{1}{c|}{1.36} & \underline{28.14} \\ \hline

\multicolumn{9}{c}{discrete tokens (autoregressive / masked transformer based)}                                                                                                                                                                                          \\ \hline
\multicolumn{1}{c|}{AudioGen$^+$~\citep{audiogen}}      & \multicolumn{1}{c|}{1.6B}        & \multicolumn{1}{c|}{101K}           & \multicolumn{1}{c|}{11s}         & \multicolumn{1}{c|}{4.09} & \multicolumn{1}{c|}{29.65} & \multicolumn{1}{c|}{7.86}  & \multicolumn{1}{c|}{1.95} & 13.24 \\ \hline
\multicolumn{1}{c|}{MagNet$^*$~\citep{ziv2024masked}}        & \multicolumn{1}{c|}{1.5B}        & \multicolumn{1}{c|}{-}              & \multicolumn{1}{c|}{4s}          & \multicolumn{1}{c|}{3.64} & \multicolumn{1}{c|}{26.11} & \multicolumn{1}{c|}{8.58}  & \multicolumn{1}{c|}{1.88} & 10.54 \\ \hline
\multicolumn{1}{c|}{DelayPattern$^*$~\citep{copet2024simple}}  & \multicolumn{1}{c|}{1.5B}        & \multicolumn{1}{c|}{-}              & \multicolumn{1}{c|}{11s}         & \multicolumn{1}{c|}{2.51} & \multicolumn{1}{c|}{12.47} & \multicolumn{1}{c|}{10.62} & \multicolumn{1}{c|}{1.96} & 14.23 \\ \hline
\multicolumn{1}{c|}{Siren (ours)}  & \multicolumn{1}{c|}{1.6B}        & \multicolumn{1}{c|}{100K+1K}        & \multicolumn{1}{c|}{13s}            & \multicolumn{1}{c|}{1.35} & \multicolumn{1}{c|}{\underline{10.65}}  & \multicolumn{1}{c|}{{12.85}} & \multicolumn{1}{c|}{{1.33}} & 24.18 \\ \hline
\multicolumn{1}{c|}{DelayPattern$^+$~\citep{copet2024simple}}  & \multicolumn{1}{c|}{3.3B}        & \multicolumn{1}{c|}{437K}           & \multicolumn{1}{c|}{25s}         & \multicolumn{1}{c|}{2.20} & \multicolumn{1}{c|}{12.50} & \multicolumn{1}{c|}{11.66} & \multicolumn{1}{c|}{1.70} & 16.58 \\ \hline
\multicolumn{1}{c|}{Siren (ours)}  & \multicolumn{1}{c|}{3.1B}        & \multicolumn{1}{c|}{436K+1K}        & \multicolumn{1}{c|}{25s}            & \multicolumn{1}{c|}{\underline{1.28}}     & \multicolumn{1}{c|}{\textbf{10.35}}      & \multicolumn{1}{c|}{\underline{13.93}}      & \multicolumn{1}{c|}{1.36}     & 25.64      \\ \whline
\end{tabular}
}}
\caption{Main results on AudioCaps test-set. \emph{\#Gen. Time} means the cost time to conduct inference with a batch size of 16 (effective sample count of 8 when using CFG) using NVIDIA-L20. The best performed metric is in \textbf{bold}, and the second best is \underline{underlined}. $^*$ means we use officially available models, while $^+$ means we train them from scratch due to unavailable models.}
\label{tab-main_result}
\end{table*}

\noindent divergence, Inception Score (IS), and CLAP score. Please see appendix for details in computation.
\subsection{Main Results}
We compare \method~with both diffusion- and LMs-based T2A generators in Table~\ref{tab-main_result}.
For fair comparison, we report model size, training data scale, latency per batch, and employ CLAP-score-based reject sampling following \citealp{GenAU}.

As shown in Table~\ref{tab-main_result}, conventional LM-based models underperform diffusion counterparts by wide margins (e.g., 2.20 vs. 1.22 FAD) despite comparable parameter counts, underscoring their struggle to leverage RVQ’s potential. In contrast, \method~achieves state-of-the-art fidelity by deploying a deep RVQ tokenizer ($r=12$) while maintaining efficient autoregressive decoding. This narrows the gap with diffusion models, demonstrating that discrete token modeling can rival continuous-domain approaches.
Notably, \method~trails MMAudio \citep{MMAudio} and AudioX \citep{audioX} in CLAP score by 2–4 points. We attribute this to their multi-modal training (text, video), which enriches semantic grounding even when inferring with text-only prompts. 
\begin{table}[htbp]
\centering
\renewcommand{\arraystretch}{1.35}
\resizebox{.5\textwidth}{!}{
\begin{tabular}{c|c|c|c|c|c|c|c|c}
\whline
\# para. & Embed. & Dim  & Main Layer & Main Para. & Decoder Layer & Decoder Para. & Head Count & Head Para. \\ \hline
1,632M   & 12.5 M & 1024 & 14         & 231M       & 2             & 25M           & 12         & 12.6       \\ \hline
3,084M   & 12.5M  & 1024 & 24         & 396M       & 8             & 103M          & 12         & 12.6       \\ \whline
\end{tabular}
}
\caption{Details of transformer' s architecture.}
\label{tab:architecture}
\end{table}
\paragraph{Architecture Details}
In this section, we present a comprehensive overview of Siren’s architectural blueprint, detailing each core component that constitutes the full generative pipeline. Our design comprises four principal modules: (1) Multimodal Embedding Layers, responsible for encoding heterogeneous input modalities (e.g., text, audio tokens, or conditioning signals) into a unified latent space; (2) Main Transformer Blocks, which form the backbone of the model and perform deep contextual reasoning over the embedded sequences; (3) Decoder Transformers, specialized modules that progressively refine latent representations into structured output tokens; and (4) Prediction Heads, lightweight output layers that map the decoder’s final representations to target audio token distributions or waveform parameters.

\subsection{Ablation Study}
\begin{table}[htbp]
\centering
\renewcommand{\arraystretch}{1.35}
\resizebox{.4\textwidth}{!}{
\begin{tabular}{c|c|c|c|c}
\whline
Setting        & \#Parameter & FAD  & FD    & CLAP  \\ \whline
Small Single   & 530M        & 8.22 & 40.44 & 4.81  \\ \hline
Base Single    & 1.6B        & 3.88 & 25.18 & 7.79  \\ \hline
Large Single   & 3.3B        & 1.92 & 11.12 & 15.64 \\ \hline
Isolated-Small & 1.6B        & 1.44 & 12.09 & 17.07 \\ \hline
Isolated-Large & 3.1B        & 1.21 & 9.90  & 20.24 \\ \whline
\end{tabular}
}
\caption{Ablating different architecture options.}
\label{tab:scale_single}
\end{table}
\textbf{Ablating among architectures}
In this subsection, we compare different architecture options, including small single LM, larger single LM, small isolated transformers, and larger isolated transformers. To conduct a fair comparison, we benchmarked our parallel architecture against a series of single Transformer models whose total parameter counts closely match those of our multi-LM setup. Crucially, the only architectural difference in this comparison is the isolation: ours uses separate models, while the baseline uses a single transformer.

The results in Table~\ref{tab:scale_single} demonstrate that our isolated LLM architecture outperforms the single-model baseline across key evaluation metrics. This suggests that the parallel, modular design itself—not just increased scale—provides a fundamental advantage, likely due to enhanced specialization or reduced interference between tasks.

\begin{table}[]
\centering
\renewcommand{\arraystretch}{1.35}{
\resizebox{.45\textwidth}{!}{
\begin{tabular}{c|c|c|c|ccc}
\whline
Collab. & Residual Con. & Accumulated Cond. & RL Align & \multicolumn{1}{c|}{FD}                           & \multicolumn{1}{c|}{ISC}                           & CLAP  \\ \whline
\ding{55}       & \ding{55}             & \ding{55}                 & \ding{55}        & \multicolumn{1}{c|}{29.65}                        & \multicolumn{1}{c|}{7.86}                          & 13.24 \\ \hline
\ding{51}       & \ding{55}             & \ding{55}                 & \ding{55}        & \multicolumn{3}{c}{\emph{itg}}                                                                                        \\ \hline
\ding{51}       & \ding{51}             & \ding{55}                 & \ding{55}        & \multicolumn{1}{c|}{14.66}                        & \multicolumn{1}{c|}{8.90}                          & 15.09 \\ \hline
\ding{51}       & \ding{55}             & \ding{51}                 & \ding{55}        & \multicolumn{1}{c|}{12.09}                        & \multicolumn{1}{c|}{10.70}                         & 17.07 \\ \hline
\rowcolor[HTML]{EFEFEF} 
\ding{51}       & \ding{55}   & \ding{51}                 & \ding{51}        & \multicolumn{1}{c|}{\cellcolor[HTML]{EFEFEF}10.65} & \multicolumn{1}{c|}{\cellcolor[HTML]{EFEFEF}12.85} & 24.18 \\ \whline
\end{tabular}
}}
\caption{Ablating among major modules.}
\label{tab-module_ablation}
\end{table}
\textbf{Ablating Method Module.} 

To validate \method's effectiveness, we present results in Table~\ref{tab-module_ablation}. Row 1 shows a baseline without specific design, yielding suboptimal performance. Introducing collaboration in row 2 allows models to fit their respective codes during training, but disrupts RVQ's causal relationship, thus inability to generation (\emph{itg}). Conditions are added to resolve this, using either the previous layer's code (residual approach) or the sum of all previous codes (accumulated approach). The latter in row 4 proves superior, so it's adopted. To reduce exposure bias, RL Alignment is added in row 5, greatly enhancing overall performance.

\begin{table}[]
\centering
\renewcommand{\arraystretch}{1.35}{
\resizebox{.3\textwidth}{!}{
\begin{tabular}{c|c|c|c}
\whline
RVQ Prediction       & Loss Mean & Loss Ratio & FAD    \\ \whline
Indpdt.              & 1.19      & 1.95       & 4.09 \\ \hline
AR                   & 0.88      & 2.17       & 4.19 \\ \hline
Accumu. AR           & 0.80      & 2.15       & 3.88 \\ \hline
\rowcolor[HTML]{EFEFEF} 
Collaborative (ours) & 0.29      & 1.55       & 1.44 \\ \whline
\end{tabular}
}}
\caption{Ablating different conditional strategies to model the causal relationship of inter-RVQ layer codes.}
\label{tab-ablate_RVQ}
\vspace{-3mm}
\end{table}

\begin{table}[]
\centering
\renewcommand{\arraystretch}{1.35}{
\resizebox{.4\textwidth}{!}{

\begin{tabular}{c|c|c|c|c|c}
\whline
Model Distribution & \#Single Para. & FD    & ISC   & CLAP  & Loss Ratio \\ \whline
1-12               & 1.6B           & 25.18 & 7.79  & 11.43 & 2.15       \\ \hline
2-6                & 800M           & 23.89 & 6.62  & 10.84 & 2.09       \\ \hline
4-3                & 400M           & 15.90 & 11.03 & 13.24 & 1.74       \\ \hline
\rowcolor[HTML]{EFEFEF} 
6-2                & 270M           & 12.09 & 10.70 & 17.07 & 1.55       \\ \hline
12-1                & 140M           & \multicolumn{4}{c}{\emph{itg}}        \\ \whline
\end{tabular}
}}
\caption{Ablating among different model distributions.}
\label{tab-ablate_model_distribution}
\end{table}
\noindent\textbf{Ablating the Condition of RVQ Prediction} 

In Table~\ref{tab-ablate_RVQ}, we focus on two metrics to further verify inter-model conditioning: Loss Mean, indicating training fit by the smallest average of RVQ layer code losses; and Loss Ratio, indicating training balance by the ratio of the $12^{th}$
 to the $1^{st}$
  code loss. Independent heads (row 1) struggle to converge, with deeper layers facing more challenges. Incorporating conventional AR mode (row 2) enhances fit but makes deeper code learning harder. Accumulation mode (row 3) conditions on rich semantic cues, greatly improving fit and balance. Combined with collaborative method (row 4) achieves excellent code balance and overall fit.
  
\begin{figure}
    \centering
    \includegraphics[width=0.5\linewidth]{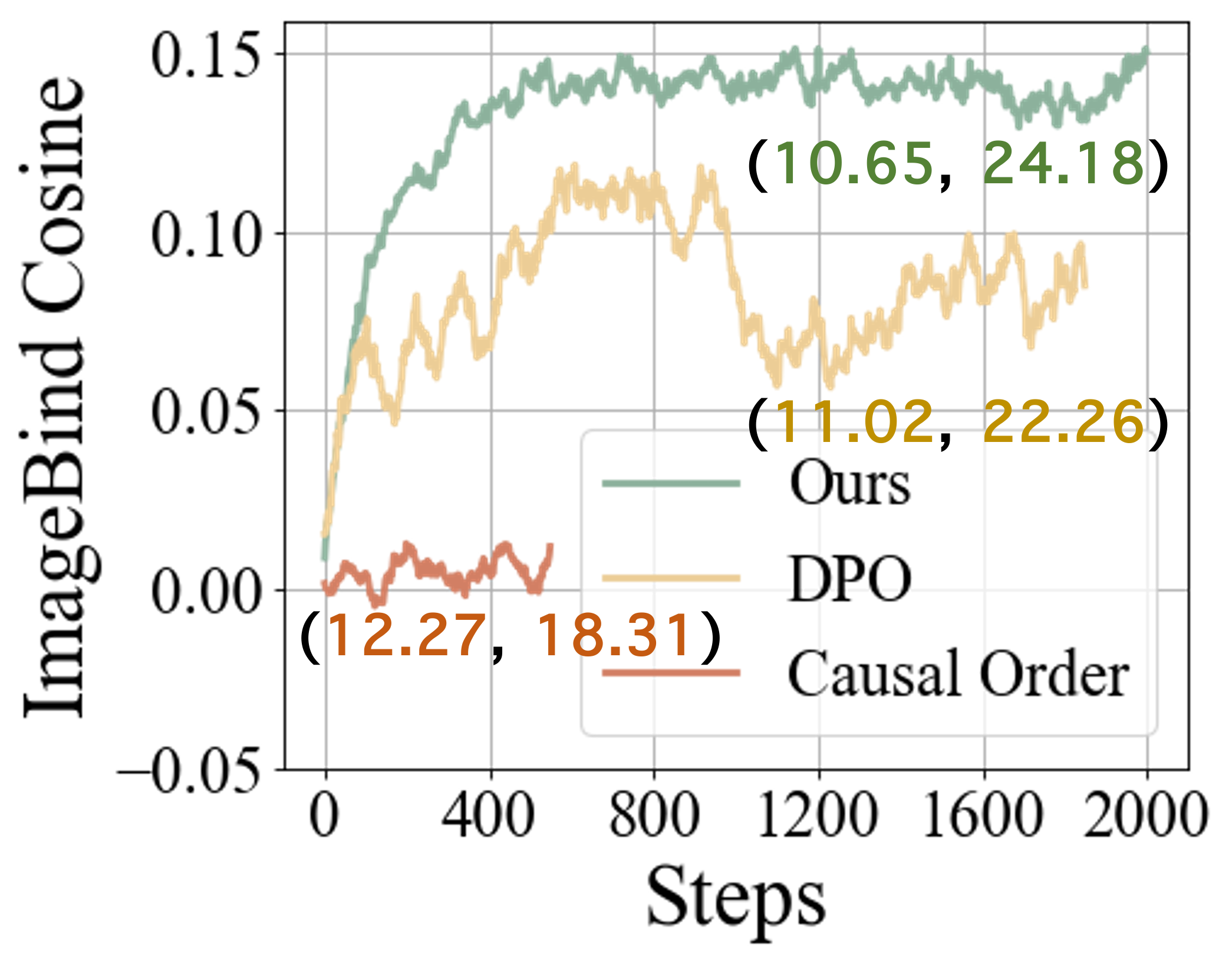}
    \caption{Rewards (ImageBind Cosine) changing curves among different reinforcement learning strategy. The figure pair aside each curve means (FD, CLAP Score).}
    \label{fig-RL_curve}
\end{figure}
\begin{table}[htbp]
\centering
\renewcommand{\arraystretch}{1.35}{
\resizebox{0.5\textwidth}{!}{
\begin{tabular}{cc|c|c|c|c}
\whline
\multicolumn{1}{c|}{Align Direction}                     & Num. Model & \# Train Parameter & CLAP  & ISC   & FD    \\ \whline
\rowcolor[HTML]{EFEFEF} 
\multicolumn{1}{c|}{\cellcolor[HTML]{EFEFEF}Anti-Causal} & 1          & 270M               & 24.18 & 12.85 & 10.65 \\ \hline
\multicolumn{1}{c|}{Anti-Causal}                         & 2          & 540M               & 25.12 & 12.77 & 10.80 \\ \hline
\multicolumn{1}{c|}{Causal}                              & 1          & 270M               & 18.31 & 9.55  & 12.27 \\ \whline
\end{tabular}
}}
\caption{Ablating reinforcement learning components.}
\label{tab-ablate_RL}
\end{table}
\noindent\textbf{Ablating the Number of RVQ Codes Per Model} 
To determine the number of codes each model handles, we present results for five variants in the Table~\ref{tab-ablate_model_distribution}. For efficiency, more models mean fewer parameters per model. For performance, the 1-12 shows significant degradation, as a single model cannot handle the diverse code distribution, and deeper layers increase prediction difficulty, seen in the larger Loss Ratio. Reducing codes per model improves focus and performance, with the 6-2 configuration being optimal. Notably, 12-1 fails to generate reasonable audio because its single model capacity is too limited to handle all temporal cues.

\noindent\textbf{Ablating the Anti-Causal Alignment} We perform ablations on the optimization direction and the number of models optimized in anti-causal alignment. Firstly, a causal direction, \emph{i.e.}, training the last model to align with earlier models' preferences, results in reward around zero in Figure~\ref{fig-RL_curve} and Table~\ref{tab-ablate_RL} row 3 confirms ineffectiveness. This is because the last two codes contain the least semantics, thus having minimal impact on the performance. Secondly, training the first two models under the anti-causal strategy achieves comparable results as shown in row 2, suggesting one model is sufficient. Besides, combining this strategy with DPO~\citep{dpo} also yields improvements, but not as effectively as our RL method, as Figure~\ref{fig-RL_curve} depicts.

\begin{table}[htbp]
\centering
\renewcommand{\arraystretch}{1.35}
\resizebox{.5\textwidth}{!}{
\begin{tabular}{c|c|c|c|c|c|c}
\whline
Method       & General Fidelity & ME. Fidelity & OOD Fidelity & General Ins. & ME. Ins. & OOD Ins. \\ \whline
DelayPattern & 27.3             & 25.4         & 29.3         & 15.2         & 17.5     & 28.3     \\ \hline
AudioX       & 28.1             & 35.9         & 33.5         & 26.5         & 39.2     & 33.8     \\ \hline
Ours         & 44.6             & 38.7         & 37.2         & 58.3         & 43.3     & 37.9     \\ \whline
\end{tabular}
}
\caption{Results from user study.}
\label{tab:user}
\end{table}
\subsection{User Study}
To better validation, we conducted a comprehensive user study involving 20 human experts, who evaluated audio samples generated by three distinct systems: AudioX (representing the state-of-the-art in diffusion models), DelayPattern (the leading language model-based approach), and our proposed method, Siren. The results have been exhibited in Table~\ref{tab:user}. In each evaluation round, raters were presented with a set of three audio clips—one from each system—and were asked to rank them according to two key dimensions: Fidelity Quality, which assesses overall audio realism and sound clarity, and Instruction Following, which measures how accurately the audio reflects the given prompt.

To further probe the models’ capabilities under more challenging conditions, we curated three specialized subsets of prompts. The first, General Scenes, consists of 30 randomly selected prompts from the AudioCaps dataset to establish a baseline. The second, Multi-Event, includes 30 prompts from AudioSet that describe scenes with more than two concurrent sound events—selected with the assistance of the Qwen3-235B-A22B~\cite{yang2025qwen3} LLM to ensure complexity. The third, OOD Prompts, comprises 30 synthetically generated prompts, also crafted using Qwen3-235B-A22B, that depict rare or unusual acoustic scenarios designed to test the models’ generalization and compositional reasoning by combining unlikely or previously unseen events.

The results, measured by the percentage of times each method was ranked best across all evaluations, show that Siren outperforms AudioX and DelayPattern in fidelity and instruction following.

\section{Related Work}
\noindent\textbf{Diffusion-Based Audio Generation}
Denoising diffusion models~\citep{ddpm,scorebased,ddim,adm,dpm-solver,ldm,dit} excel in continuous-domain generation by iteratively refining noisy signals into structured outputs, making them a natural fit for audio synthesis. Prior work applies these models to audio via latent-space diffusion: 1D tokenizer operate on waveform embeddings~\citep{SAO,ETTA}, while 2D architectures process mel-spectrograms using U-Nets~\citep{liu2022diffsinger,liu2023audioldm,liu2024audioldm,xue2024auffusion,evans2024long,xing2024seeing,du2023conditional,liu2024tell,agostinelli2023musiclm,Tango2,MelQCD} or diffusion transformers (DiT)~\citep{ETTA,Fugatto}. Recent advance~\citep{MMAudio,Fugatto} integrate flow matching~\citep{FlowMatching,FMRecon} to accelerate sampling, achieving state-of-the-art fidelity. 
However, reliance on continuous latent representations creates a fundamental divergence from discrete token-based paradigms like LMs, complicating efforts toward unified multi-modal frameworks~\citep{wu2023next,fei-etal-2024-empathyear,QWen25_Omni,Crab}.\\
\noindent\textbf{Language Models for Audio Synthesis}
Inspired by successes in NLP, LM-based audio generators map waveforms to discrete tokens via vector quantization (VQ)~\citep{igpt,maskgit,mage,magvit,movq,muse,var}. Early approaches paired VAEs with LMs, but VQ’s lossy compression limited audio fidelity~\citep{magvit2,fsq}. Residual VQ (RVQ)~\citep{wu2019vector,encodec,dac} emerged as a dominant alternative. While RVQ-enhanced LMs improve fidelity, their performance remains sensitive to the choice of RVQ layers ($r$): deeper hierarchies strain autoregressive modeling due to feature orthogonality and exposure bias. Our work directly addresses these limitations through architectural innovations that disentangle RVQ layer-specific learning while preserving cross-layer coherence.
\section{Conclusion}

To enable LMs to predict multi-RVQ codes with a deep RVQ layer in text-to-audio generation, we propose \method, a novel framework that employs collaborative transformers with anti-causal alignment. 
By disentangling RVQ code-specific conditional learning objectives and harmonizing cross-model conditional reference alignment via reinforcement learning, \method~achieves SOTA performance.

\section*{Limitations}
While \method~advances autoregressive text-to-audio generation, three limitations merit discussion:
\underline{(1) Training Efficiency}: Partitioning $r$ RVQ codes across $r/2$ isolated transformers mitigates gradient conflicts and enables training billion-parameter models (e.g., 1.6B–3.1B) on consumer-grade GPUs (e.g., lower to 24GB VRAM). However, this design increases wall-clock training time: deploying \method~requires sequential training of up to six transformer modules without sufficient parallelization when no available devices. Future work will explore hybrid RVQ tokenizers that achieve comparable reconstruction fidelity with fewer layers ($r$), reducing both model number and training overhead.
\underline{(2) Model Size vs. Semantic Richness}: Although \method~matches diffusion models in inference speed and surpasses them in fidelity metrics (e.g., FD), its parameter count exceeds diffusion counterparts. We attribute this to the inherent trade-off between RVQ’s discrete tokens (low semantic density) and LM scalability: richer semantics per token could enable smaller models. Improving tokenizers to encode higher-level acoustic semantics—akin to linguistic units in text—remains a critical direction.
\underline{(3) Data Scaling}: Our experiments use a curated dataset smaller than those in prior work (e.g., GenAU, MMAudio). While rigorous filtering ensures quality, expanding data diversity—particularly with multi-modal (text, video) or long-form audio—could enhance semantic grounding and temporal coherence. Future efforts will prioritize scalable data collection pipelines to balance quality and quantity.
\section*{Acknowledgement} 
This work was partially supported by RGC Collaborative Research Fund (No. C5055-24G), the Start-up Fund of The Hong Kong Polytechnic University (No. P0045999), the Seed Fund of the Research Institute for Smart Ageing (No. P0050946), and Tsinghua-PolyU Joint Research Initiative Fund (No. P0056509), and PolyU UGC funding (No. P0053716). 

\bibliography{ref}
\clearpage
\appendix

\section{Data Engine}
\label{app:data engine}
We detail our data pipeline for curating high-quality audio-text pairs, comprising four stages: collection, preprocessing, captioning, and filtering.
\paragraph{Data Collection}
Following established practices in T2A research~\citep{xue2024auffusion,GenAU}, we aggregate audio from public repositories. For video-derived audio, we extract and segment raw audio tracks, retaining only segments with detectable acoustic activity (SNR > 6dB). This yields \textbf{2.1 million unlabeled audio clips} spanning 2–100 seconds.
\paragraph{Preprocessing}
To align inputs with our RVQ tokenizer~\citep{dac}, we standardize waveforms as follows:
Channel Conversion: Convert stereo/multi-channel audio to mono.
Resampling: Downsample to 16 kHz to match the tokenizer’s Nyquist frequency.
Segmentation: Split clips exceeding 10 seconds into fixed-length 10s chunks using a sliding window (stride=10s, no overlap), discarding residual segments (<10s). Shorter clips are zero-padded to 10s.
This produces \textbf{4.2 million standardized 10s audio segments}, doubling the initial dataset through segmentation.
\paragraph{Captioning}
Only data from AudioCaps, Clotho include human-annotated captions. For the remaining, we generate synthetic captions using QWen-2 Audio, a multimodal LLM fine-tuned for acoustic understanding. Each audio segment receives five candidate captions via the prompt:
\begin{quote}
\texttt{Please describe this audio in detail, including events, timbres, temporal structure, and emotional tone.}
\end{quote}
To refine captions, we apply Qwen-2.5 LLM with rule-based filtering:
Removal: Eliminate non-descriptive text (e.g., transcribed speech, non-English content).
Normalization: Standardize acoustic terms (e.g., “barking” → “dog bark,” “low-pitched rumble” → “engine noise”).
This ensures captions are concise, objective, and acoustically grounded.
\paragraph{Data Filtering}
We compute audio-text alignment scores using CLAP~, retaining pairs where:
\begin{equation*}
\text{CLAP-score} = \frac{\text{Audio} \cdot \text{Text}}{|\text{Audio}| |\text{Text}|} \geq \tau
\end{equation*}
For the 1.6B \method~model, we set 
$\tau=0.4$ yielding 117k high-confidence pairs. From these, we:

Reserve the top 1k (highest CLAP-score) for reinforcement learning (RL) fine-tuning.
Randomly select 100k for base model pretraining.
To scale to the 3.1B model, we lower $\tau$ to 0.35, adding 320k moderately aligned pairs. Combined with the initial 116k, this forms a 436k training corpus, balancing quality and diversity.

\section{Experimental Details}
\label{app: exp details}
\paragraph{Embedding Layer Setup}
The RVQ tokenizer from~\citealp{dac} employs low-dimensional codebooks (dim=8) followed by a post-projector to restore waveform fidelity. To adapt these tokens for transformer-based generation, we retain all quantization layers and introduce 12 parallel MLP projectors to map discrete token indices into high-dimensional embeddings compatible with the transformer’s hidden dimension. This preserves quantization fidelity while ensuring seamless integration with the autoregressive architecture.
\paragraph{Textual Condition Encoder}
Text prompts are encoded into embeddings via the frozen CLAP-Text encoder~\citep{CLAP}. These embeddings are injected into the transformer through cross-attention layers in all decoder blocks, where CLAP embeddings serve as keys/values and audio tokens as queries. 
\paragraph{Training Protocol}
This training employs the AdamW optimizer with a 3e-4 learning rate and 24 batch size on each NVIDIA L20 GPU, which costs around $600$ GPU hours for $1.6B$ variant, and $2.5K$ for $3.1B$ variant.
To conduct data augmentation, we randomly crop 6s segments from 10s audio, retaining only crops with CLAP similarity higher than 0.20 to their original caption.

\paragraph{Inference \& Sampling}
Autoregressive decoding is accelerated using KV-caching. While classifier-free guidance (CFG) is common in generative models, its integration with reinforcement learning-based anti-causal alignment remains unstable; thus, we disable CFG. Bridging this gap—enabling controllable generation via guidance in LM-based systems—is a key direction for future work.
\section{Additional Technical Details}
\subsection{Metrics}
The main objective evaluation metrics we use are Frechet Distance (FD), Inception Score (IS), and Kullback–Leibler (KL) divergence. These metrics are based on the state-of-the-art audio classifier PANNs~\cite{kong2020panns}. Specifically, FD, analogous to the Frechet Inception Distance in image generation, measures the similarity between generated and target audio samples. IS evaluates both sample quality and diversity. KL divergence is computed at the paired-sample level and averaged for the final score. We also report Frechet Audio Distance (FAD)~\cite{kilgour2018fr}, which follows a similar principle but uses VGGish~\cite{hershey2017cnn}—a potentially less effective classifier than PANNs. In addition, the CLAP score~\cite{CLAP} is adopted to measure the semantic alignment of audio-text.
\subsection{Tokenizer}
\label{app:tokenizer}
Consider an audio waveform $x\in\mathbb{R}^{l_{wav}\times C_{wav}}$, where $C_{wav}$ is the number of channel, $l_{wav}$ is the duration length.
The goal of tokenization is to compress $x$ into the latent space $f\in \mathbb{R}^{l\times C}$, where empirically we have $C > C_{wav}$, and $l < l_{wav}$.
To apply autoregressive transformer modeling to audios via next-token prediction, we must tokenize an audio into $l$ \emph{discrete} tokens. 
\paragraph{Vector Quantization (VQ)}
To this end, VQ firstly converts an audio into continuous tokens (feature) $f\in \mathbb{R}^{l\times C}$, which is then quantized into discrete tokens $q\in[V]^l$:
\begin{equation}
    f=\mathcal{E}(x),\qquad q=\mathcal{Q}(f),
\end{equation}
where $\mathcal{E}$ denotes a tokenizer, $\mathcal{Q}$ a quantizer. 
The quantizer typically includes a learnable codebook $Z\in\mathbb{R}^{V\times C}$ containing $V$ vectors.
The quantization process can be described as mapping each entry $t$ ($t\le l$) of feature $f_{t}$ to the code index $q_{t}$ of its nearest code in the Euclidean sense:
\begin{equation}
    q_{t}=(\mathop{\arg\min}_{v\in[V]} \ \ 
    \| 
        \texttt{lookup}(Z, v)-f_t
    \|_2)\in [V],
\end{equation}
where $\texttt{lookup}(Z,v)$ denotes taking the $v^{th}$ vector in codebook $Z$.
Then, $Z$ is looked up by every entry $q_t$ independently to retrieve corresponding code $\hat{f}_{t}\in \mathbb{R}^{1\times C}$, which is an approximate of original feature $f_{t}$.
To recover the $\hat{f}$ into soundable waveform, a de-tokenizer $\mathcal{D}$ is applied, where the whole modules are trained with an audio reconstruction goal:
\begin{equation}
 \hat{x}=\mathcal{D}(\hat{f}),\qquad \mathop{\arg\min}_{\mathcal{E},\mathcal{Q},\mathcal{D}}\mathcal{L}_{\mathtt{Recon}}(\hat{x}, x),
\end{equation}
where $\hat{x}$ is the reconstructed version of audio $x$, and $\mathcal{L}_{\mathtt{Recon}}$ is a general representation, with different designs in existing works.

\paragraph{Audio Generation by Autoregressive Transformer}
With an audio $x$ represented as discrete, we can derive a sequence of codes $\{q_t|t=1,2,...,l; 0\le q_t < V\}$. 
Then, the autoregressive transformer produces a distribution over codebook by a $V$-way classifier to fit a conditional distribution of $p(q_t|q_{<t},c)$, where $c$ can be any conditions to control the generated audio, like textual prompts in this paper.
During inference, commencing from a user given $c$, the transformer $\mathcal{F}$ predicts next time-step discrete token autoregressively.
After obtained all of tokens, they are concatenated along the temporal dimension, then utilized to lookup the codebook to derive the codes, which are then feed into de-tokenizer to derive generated waveform $\tilde{x}$.

\subsection{Algorithms of Siren}
\label{app:algorithms}

\begin{algorithm}[H]
\caption{First Stage Training Pipeline of \method}
\begin{algorithmic}[1]
\Require Ground-Truth RVQ tokens $\mathbf{q}\in[V]^{r\times l}$;
\Require Hyperparameters: the number of RVQ layer $r$, the number of transformer model $K=r/2$;
\Require Each transformer $\mathcal{H}_k$ is composed of backbone model $\mathcal{F}_k$, residual layer decoding model $\mathcal{R}_k$, and two classifiers $(\mathcal{C}_{2k},\mathcal{C}_{2k+1})$.
\For{$k = 1, \cdots, K$}
    \For{train loops}
        \State \emph{Predict next-time feature}: $(h_{1\cdots l})_k = \mathcal{F}(\sum_{j=1}^r\texttt{lookup}(Z^j,\mathbf{q}_{sos,1\cdots l-1}^j),c)$  ;
        \State \emph{Predict next-r-codes}:
        $\hat{\mathbf{q}}^{2k}_{1\cdots l}=\mathcal{C}_{2k}\mathcal{R}_k((h_{1\cdots l})_k, \mathbf{q}^{<2k}_{1,\cdots,l})$;
        \State \emph{Predict next-r-codes}:
        $\hat{\mathbf{q}}^{2k+1}_{1\cdots l}=\mathcal{C}_{2k+1}\mathcal{R}_k((h_{1\cdots l})_k, \mathbf{q}^{<2k+1}_{1,\cdots,l})$;
        \State \emph{Compute loss}: $\mathcal{L}=\mathcal{L}_{ce}(\hat{\mathbf{q}}^{2k}_{1\cdots l},\mathbf{q}^{2k}_{1\cdots l})+\mathcal{L}_{ce}(\hat{\mathbf{q}}^{2k+1}_{1\cdots l},\mathbf{q}^{2k+1}_{1\cdots l})$;
    \EndFor
\EndFor
\Ensure $\{\mathcal{H}_k|\mathcal{H}_k=\mathcal{F}_k\times\mathcal{R}_k\times\mathcal{C}_{2k} \times\mathcal{C}_{2k+1} \}_{k=1}^{r/2}$
\end{algorithmic}
\end{algorithm}

\begin{algorithm}[H]
\caption{Sampling Pipeline of \method}
\begin{algorithmic}[1]
\Require Trained $r/2$ transformers in the first/second stage;
\Require Detokenizer $\mathcal{D}$;
\For{$t=1,\cdots,l$}
    \For{$k=1,\cdots,r/2$}
        \State \emph{Predict next-time feature}: $(h_t)_k = \mathcal{F}(\sum_{j=1}^r \texttt{lookup}(Z^j, \mathbf{q}^j_{sos,<t}), c)$;
        \State \emph{Predict next-first-code}: $\hat{\mathbf{q}}_t^{2k} = \mathcal{C}_{2k}(\mathcal{R}_k((h_t)_k, \hat{\mathbf{q}}_t^{<2k}))$;
        \State \emph{Predict next-second-code}: $\hat{\mathbf{q}}_t^{2k+1} = \mathcal{C}_{2k+1}(\mathcal{R}_k((h_t)_k, \hat{\mathbf{q}}_t^{<2k+1}))$;
        \State \emph{Update}: $\hat{\mathbf{q}} \gets \hat{\mathbf{q}}_t^{2k}$, $\hat{\mathbf{q}} \gets \hat{\mathbf{q}}_t^{2k+1}$.
    \EndFor
\EndFor
\State $\hat{x} = \mathcal{D}(\hat{\mathbf{q}})$
\Ensure $\hat{x}$
\end{algorithmic}
\end{algorithm}

\begin{algorithm}[H]
\caption{Second Stage Training Pipeline of \method}
\begin{algorithmic}[1]
\Require Trained $r/2$ transformers in the first stage;
\Require Detokenizer $\mathcal{D}$ and ImageBind models $\phi_{audio},\phi_{text}$;
\Require Hyperparameters: Advantage threshold $\gamma$, clip thresholds $\epsilon_d,\epsilon_u$;
\For{train loops}
    \State Sample a batch of prompts $\{c_1,...,c_B\}$;
    \State Update the old policy model $(\pi_{\theta_1})_{old}\gets \pi_{\theta_1}$;
    \State Sample $G$ roll-out $\{(\hat{\mathbf{q}}_{1,...,l})_i\}_{i=1}^G\sim(\pi_{\theta_1})_{old}(\cdot|c_b)$ for each prompt $c_b,b\le B$;
    \State Compute rewards $\{R_i\}_{i=1}^G$ for each roll-out sequence by Eq.~\ref{eq:reward};
    \State For each $(\hat{\mathbf{q}}_{1,...,l})_i$ in the buffer, compute advantage $A_i$ via Eq.~\ref{eq:advantage}, and filter long $G$ using $\gamma$;
    \For{iteration=1,...,$\mu$}
    \State Update the policy model $\pi_{\theta_1}$ by maximizing the objective in Eq.~\ref{eq:RLObject}.
    \EndFor
\EndFor
\Ensure $\theta_1$
\end{algorithmic}
\end{algorithm}

\section{Additional Experimental Results}
\subsection{Property 1\& Influence 1}
\begin{figure*}
    \centering
    \includegraphics[width=0.7\linewidth]{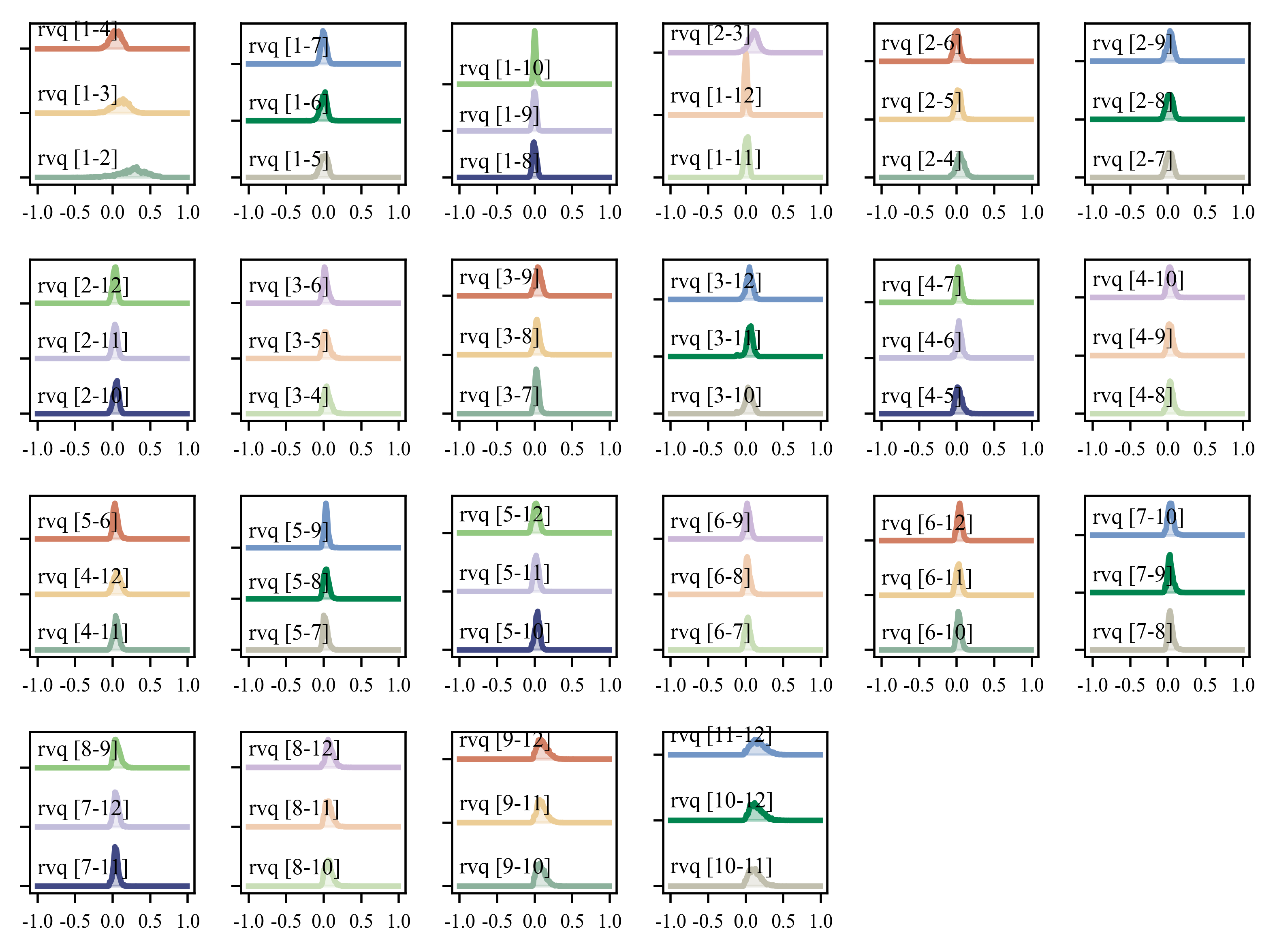}
    \caption{Cosine distributions between quantized features from any two RVQ layers.}
    \label{app_fig:Cosine}
\end{figure*}

We extract each layer' s quantized features and average pool them along temporal dimension. Hence, we derive a feature $\hat{f}^{j}=\texttt{pool}(\hat{f}^{j}_{1,...,l})$. Then, we compute the cosine similarity between quantized features from any two different layers.
As shown in Figure~\ref{app_fig:Cosine}, we illustrate the distribution of cosine similarity between any-two layers.
It is evident that the overall cosines are distributed around $0$, which represents the near orthogonality.
It is noted that the cosines between the adjacent layers from begining or ending are distributed relatively larger than 0. 
We deem that it is because of local homogeneity among those adjacent layer when represented knowledge is extreme, \emph{extremely semantic rich or poor}.
As for gradient distributions in influence 1, we gather the gradients tensor, usually in two-dimension, and average along output dimension to derive the gradient influence to the input neurons. Then, we compute any two of layers gradient vector angles and derive a $90$ degree-around distribution, which also supports our claim that orthogonality incurs diverse optimization direction 
\begin{figure}[t]
    \centering
    \includegraphics[width=\linewidth]{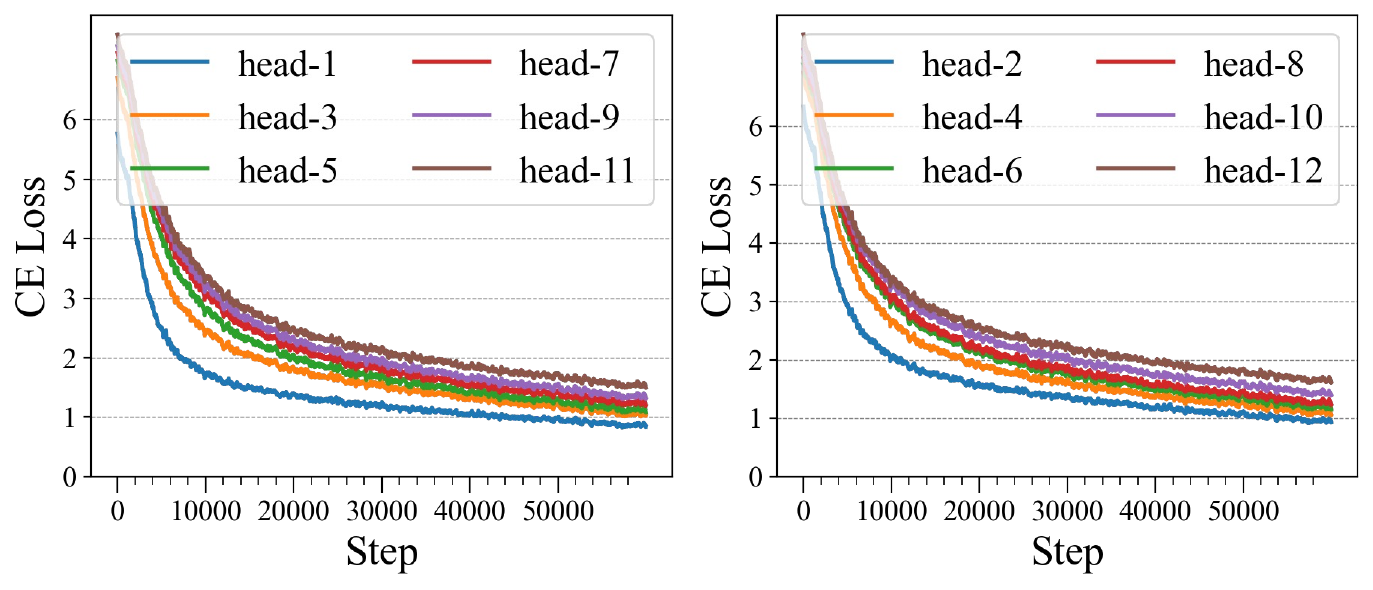}
    \caption{Training convergence curves of learning different RVQ layers.}
    \label{app_fig:loss_curve}
\end{figure}
aggregated to the input neurons.

\subsection{Property 2\& Influence 2}

\begin{figure*}
    \centering
    \includegraphics[width=0.7\linewidth]{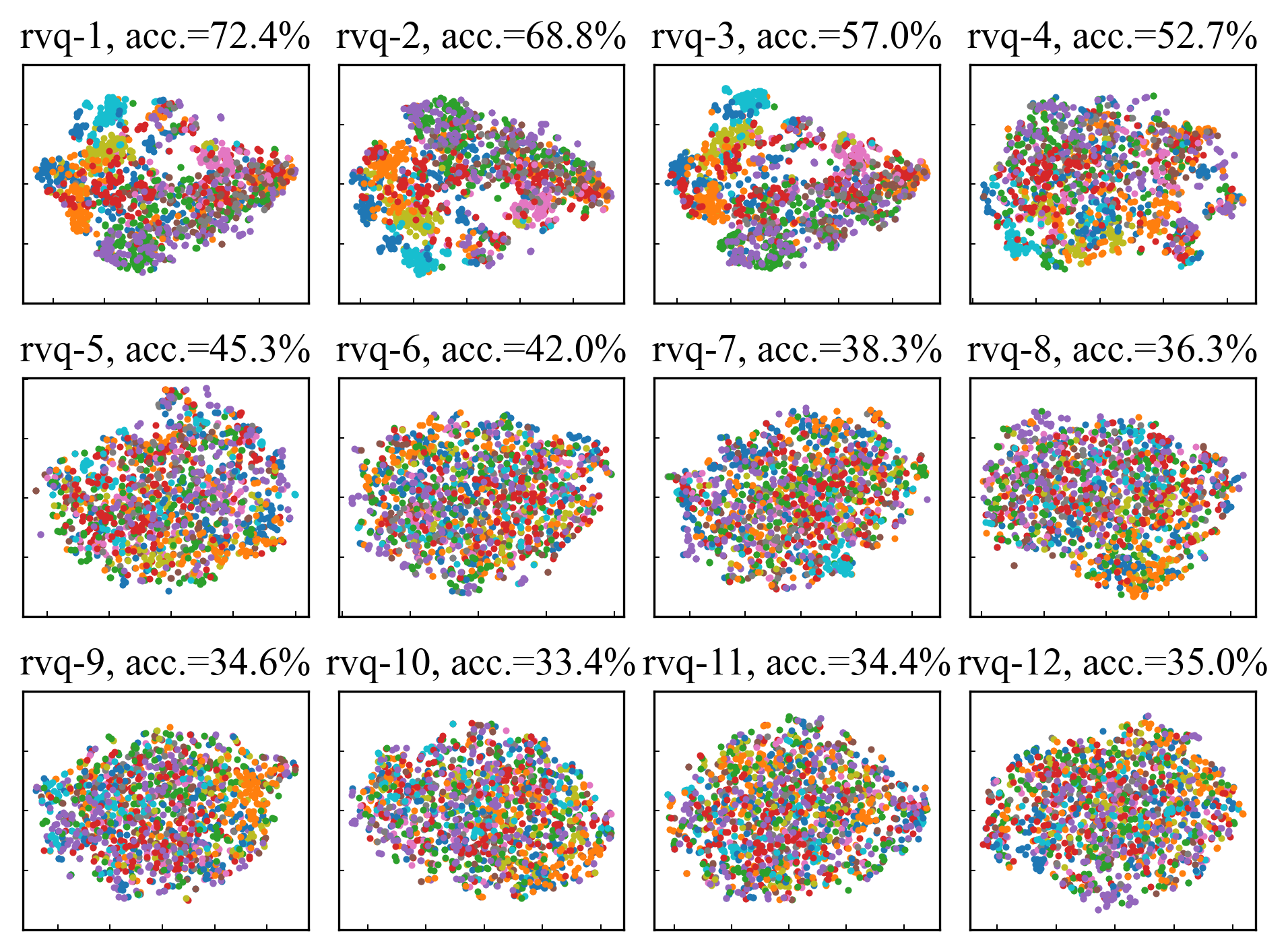}
    \caption{T-SNE processed quantized features' s distribution and corresponding top-1 classification accuracy.}
    \label{app_fig:accuracy}
\end{figure*}

In this section, we firstly complement the details of training the classifiers.
Given an audio sample, that has been tokenized into $\mathbf{q}\in[V]^{r\times l}$, we extract each row vector $\mathbf{q}^{j}\in[V]^{l}$ as input tokens, and use it to loop up corresponding codebooks to derive classifiers' input features $\hat{f}^j\in\mathbb{R}^{l\times C}$, where $C$ is the number of feature channel.
Then, we further introduce two MLP layers to conduct $15$-way classification on AVSync-15 dataset. Then, we report the top-1 accuracy over test set.
As shown by Figure~\ref{app_fig:accuracy}, the classification accuracy is descending as RVQ layer goes deeper. According the related research in representation learning~\citep{MoCo}, one can infer the semantic richness within a representation from its corresponding classification accuracy performance.
For example, we can distinguish objects according to its semantic information.
Thus, we can derive a conclusion in Property 2.
Then, we further illustrate the training curves of AudioGen on our 100K version dataset in Figure~\ref{app_fig:loss_curve}.
The convergence processes of different heads that is responsible for RVQ layers show that descending semantics incurs imbalanced learning difficulties.

\subsection{Analytic Results}
\textbf{The Necessarty of Introducing Collaboration.}
\begin{figure}
    \centering
    \includegraphics[width=\linewidth]{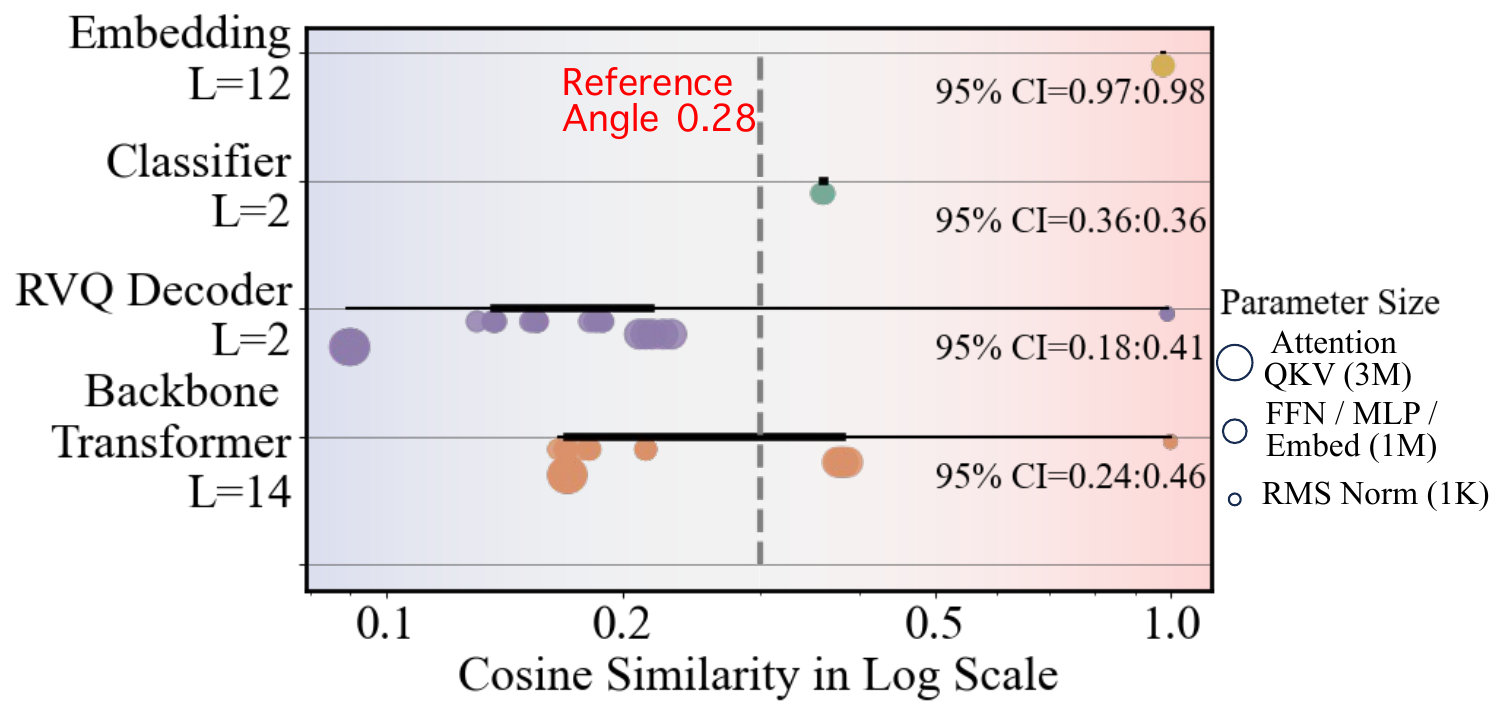}
    \caption{Cosine similarity between the same layer' s \emph{task vector} from different learned transformer. Each scatter denotes an average cosine value over all transformer' s two-combinations. The boxplots exhibit the quartiles and extremes over each module' s \emph{task vector}. As a reference, we put the average cosine values between the models in $10^{th}$ epoch with the last epoch in red fonts in this figure.}
    \label{fig:enter-label}
\end{figure}We define the task vector as the difference between the weights of the best six models and the same initialization weights, reflecting the impact of different tasks on model weights. As shown in Figure~\ref{fig:enter-label}, we categorize the model into four structures: Embedding, Classifier, RVQ Decoder, and Backbone Transformer, as indicated on the vertical axis. The RVQ Decoder and Backbone Transformer have three layers with distinct parameters, represented by circles of varying diameters. The horizontal axis shows the weight similarity among the same layers within each structure, totaling 15 combinations.

It can be observed that the RVQ Decoder establishes causal relationships between codes, striving for orthogonality between each code, thus exhibiting lower similarity. The Backbone Transformer and Classifier follow, influenced by the task and sharing some common characteristics. Other structures and layers have little relationship with the task, displaying higher similarity among them. This indicates that the core structure and the codes each model handles are closely related, suggesting the necessity for separation.


\end{document}